\documentclass[12pt,preprint]{aastex}
\tighten
\eqsecnum
\accepted{}
\journalid{}{}
\articleid{}{}
\shortauthors{ }
\shorttitle{Candidate Solar Twins} 
\received{2005 June 24}
\begin{document}

\title{Keck/HIRES Spectroscopy of Four Candidate Solar Twins}

\author{Jeremy R. King}
\affil{Department of Physics and Astronomy, 118 Kinard Laboratory,\\
Clemson University, Clemson, SC{\ \ }29634-0978}
\email{jking2@ces.clemson.edu}

\author{Ann M. Boesgaard\altaffilmark{1}}
\affil{Institute for Astronomy, 2680 Woodlawn Drive, Honolulu, HI  96822}
\email{boes@ifa.hawaii.edu}

\author{Simon C. Schuler}
\affil{Department of Physics and Astronomy, 118 Kinard Laboratory,\\
Clemson University, Clemson, SC{\ \ }29634-0978}
\email{sschule@ces.clemson.edu}

\altaffiltext{1}{Visiting Astronomer, W.M. Keck Observatory, jointly operated by the
California Institute of Technology and the University of California.}

\begin{abstract}

We use high S/N, high-resolution Keck/HIRES spectroscopy of 4 solar twin candidates (HIP 71813, 76114, 77718, 78399) 
pulled from our {\it Hipparcos\/}-based \ion{Ca}{2} H \& K survey to carry out parameter and 
abundance analyses of these objects.  Our spectroscopic $T_{\rm eff}$ estimates are some ${\sim}100$ K hotter than
the photometric scale of the recent Geneva-Copenhagen survey; several lines of evidence suggest the photometric
temperatures are too cool at solar $T_{\rm eff}$.  At the same time, our abundances for the 3 solar twin candidates
included in the Geneva-Copenhagen survey are in outstanding agreement with the photometric metallicities; there is 
no sign of the anomalously low photometric metallicities derived for some late-G UMa group and Hyades dwarfs.  A first
radial velocity determination is made for HIP 78399, and $UVW$ kinematics derived for all stars.  HIP 71813 appears
to be a kinematic member of the Wolf 630 moving group (a structure apparently reidentified in a recent analysis of late-type
{\it Hipparcos\/} stars), but its metallicity is 0.1 dex higher than the most recent estimate of this group's metallicity. 
While certainly ``solar-type'' stars, HIP 76114 and 77718 are a few percent less massive, significantly older, and metal-poor
compared to the Sun; they are neither good solar twin candidates nor solar analogs providing a look at the Sun at
some other point in its evolution.  HIP 71813 appears to be an excellent solar analog of age ${\sim}8$ Gyr.  Our
results for HIP 78399 suggest the promise of this star as a solar twin may be equivalent to the ``closest ever solar twin'' 
HR 6060; follow up study of this star is encouraged.   

\end{abstract}


\section{Introduction}

The deliberate search for and study of solar analogs has been ongoing for nearly 30 years, initiating with the seminal early
works of Hardorp (e.g., Hardorp 1978).  \citet{CdS} gives an authoritative review of this early history, many 
photometric and spectroscopic results, and the astrophysical motivations for studying solar analogs.   As of a decade ago, these 
motivations were of a strong fundamental and utilitarian nature, seeking answers to such questions as:  (a) what is the solar 
color? {\ }b) how well do photometric indices predict spectroscopic properties? {\ }c) how robust are spectral types at describing 
or predicting the totality of a stellar spectrum? {\ }d) are there other stars that can be used as exact photometric and/or 
spectroscopic proxies for the Sun in the course of astrophysical research programs? 

While these important questions remain incompletely answered and of great interest, the study of solar analogs and search
for solar twins has taken on renewed importance.   Much of this has been driven by the detection of planetary companions 
around solar-type stars; the impact of these detections on solar analog research was foreshadowed with great prescience
by \citet{CdS}.  Precision radial velocity searches for exoplanets are most robust when applied to slowly rotating and
inactive stars; solar analogs are thus fruitful targets--metal-rich ones apparently even more fruitful \citep{FV05}. 
The appeal in searching for elusive terrestrial exoplanets around solar analogs remains a natural one given the existence
of our own solar system.  

Solar analogs of various age also provide a mechanism to examine the past and future evolution of the Sun without significant or 
total recourse to stellar models.  Such efforts looking at the sun in time \citep{RGGA} now appear to be critical 
complements to studying the evolution of planets and life surrounding solar-type stars.  For example, it has been suggested
that solar-type stars may be subject to highly energetic superflare outbursts, perhaps induced by orbiting planets, 
that would have dramatic effects on atmospheres surrounding and lifeforms inhabiting orbiting planets \citep{RS2000,SKD}. 
It also seems clear that the nominal non-stochastic gradual evolution of solar-type chromospheres has important 
implications for a diversity of planetary physics (in our own solar system and others):  the structure and chemistry of 
planetary atmospheres, the water budget on Mars, and even the evolution of planetary surfaces \citep{RGGA}; such issues are 
critical ones to understand in the development and evolution of life. 

The utilitarian importance of studying solar analogs has also persisted.  For example, there should be little argument that 
differential spectroscopic analyses performed relative to the Sun are most reliable when applied to stars like the
Sun-- early G dwarfs.  Happily, such objects can be found in a large variety of stellar populations having an extreme range 
of metallicity and age.  The development of large aperture telescopes and improved instrumentation such as multi-object
spectrographs and wide field imagers over the next decade or so mean that the stellar astronomy community is poised to undertake 
abundance surveys of tens or hundreds of thousands of Galactic stars.  Critical questions confronting such ambitious but
inevitable initiatives include: {\ }a) how reliable are photometric metallicities?  b) can low-resolution spectroscopy
yield results as robust as those from high-resolution spectroscopy?  c) will automated spectroscopic analyses needed
to handle such large datasets yield reliable results?  All these questions can be addressed well by comparison with the 
results of high-resolution spectroscopy of solar analogs.   

Despite the importance of carrying out high-resolution spectroscopic analyses of solar analogs, efforts at doing so have been
deliberate in pace.  Recent exceptions to this include the solar analog studies of \citet{GHH} and \citet{ST}.  
Here, we present the first results from a small contribution aimed at remedying this pace of study.  Using the 
results of Dr.~D.~Soderblom's recent chromospheric \ion{Ca}{2} H \& K survey of nearby ($d{\le}60$ pc) late-F to early-K dwarfs
in the {\it Hipparcos\/} catalog, we have selected a sample of poorly-studied solar twin candidates having 
$0.63{\le}(B-V){\le}0.66$, \ion{Ca}{2} chromospheric fluxes within a few tenths of a dex of the mean solar value, and
$M_V$ within a few tenths of a magnitude of the solar value; there are roughly 150 such objects accessible from 
the northern hemisphere.  These objects have been or are being observed as time allows during other observing programs.  Here, 
we present echelle spectroscopy of 4 candidates obtained with Keck/HIRES.  The objects are HIP 71813, 76114, 77718, and 78399.

\section{Data and Analysis}

\subsection{Observations and Reductions}

Our 4 solar twin candidates were observed on UT July 8 2004 using the Keck I 10-m, its HIRES echelle
spectrograph, and a Tektronix $2048{\times}2048$ CCD detector.  The chosen slit width and cross-disperser
setting yielded spectra from 4475 to 6900 {\AA} at a resolution of $R{\sim}45,000$.  Exposure times ranged 
from 3 to ${\sim}6$ minutes, achieving per pixel S/N in the continuum near 6707 {\AA} of ${\sim}400$.  A log 
of the observations containing cross-identifications is presented in Table 1.  Standard reductions were 
carried out including debiasing, flat-fielding, order identification/tracing/extraction, and wavelength 
calibration (via solutions calculated for an internal Th-Ar lamp).  The H$\alpha$ and H$\beta$ features 
are located at the blue edge of their respective orders; the lack of surrounding wavelength coverage with 
which to accomplish continuum normalization thus prevented us from using Balmer profile fitting to independently
determine $T_{\rm eff}$.  Samples of the spectra in the ${\lambda}6707$ \ion{Li}{1} region can be found
later in Figures 3 and 4.   
\marginpar{Tab.~1}

\subsection{Parameters and Abundances}

Clean, ``case a'' \ion{Fe}{1} and \ion{Fe}{2} lines from the list of \citet{Th90} were selected for measurement
in our 4 solar twin candidate spectra and a similarly high S/N and $R{\sim}45,000$ Keck/HIRES lunar spectrum (described 
in \citet{King97}) used as a solar proxy spectrum.  Equivalent widths were measured using the profile fitting
routines in the 1-d spectrum analysis software package {\sf SPECTRE} \citep{FS87}.  Line strengths of all the 
features measured in each star and our solar proxy spectrum can be found in Table 2.  Abundances were derived from
the equivalent widths using the 2002 version of the LTE analysis package {\sf MOOG} and Kurucz model atmospheres 
interpolated from ATLAS9 grids.  Oscillator strengths were taken from \citet{Th90}; the accuracy of these is
irrelevant inasmuch as normalized abundances [x/H] were formed on a line-by-line basis using solar abundances
derived in the same manner.  The solar model atmosphere was characterized by $T_{\rm eff}=5777$ K, log $g=4.44$, 
a metallicity of [m/H]=0., and a microturbulent velocity of ${\xi}=1.25$; the latter is intermediate to the 
values of ${\xi}$ from the calibrations of \citet{E93} and \citet{AP2004}.  An enhancement factor of 2.2 was applied 
to the van der Waals broadening coefficients for all lines.  
\marginpar{Tab.~2}

Stellar parameters were determined as part of the Fe analysis in the usual fashion.  $T_{\rm eff}$ and ${\xi}$ were 
determined by requiring zero correlation coefficient between the {\it solar normalized\/} abundances (i.e., [Fe/H]; 
again, accomplished on a line-by-line basis) and the lower excitation potential and reduced equivalent width, 
respectively.  This approach leads to unique solutions when there is no underlying correlation between excitation 
potential and reduced equivalent width.  We show in Figure 1 that there is no such underlying correlation in our
\ion{Fe}{1} sample.  Figure 2 displays the \ion{Fe}{1}-based line-by-line [Fe/H] values versus both lower excitation
potential (top) and reduced equivalent width (bottom) using our final model atmosphere parameters for the case
of HIP 76114; the linear correlation
coefficients in both planes are ${\sim}0.00$.  Our abundance analysis is thus a purely differential one, and the derived 
parameters do not depend on the rigorous accuracy of the $gf$ values.  The 1$\sigma$ level uncertainties in $T_{\rm eff}$ 
and ${\xi}$ were determined by finding the values of these parameters where the respective correlation coefficients 
became significant at the 1${\sigma}$ confidence level.  Gravity estimates were made via ionization balance of Fe.  
The error estimates for log $g$ include uncertainties in both [\ion{Fe}{1}/H] and [\ion{Fe}{2}/H] due to measurement 
uncertainty, $T_{\rm eff}$ errors, and ${\xi}$ errors.  The final parameters and their uncertainties can be found in 
the summary of results in Table 4. 
\marginpar{Fig.~1}
\marginpar{Fig.~2}

Abundances of Al, Ca, Ti, and Ni were derived in a similar fashion using the line data in Table 2 and model
atmospheres characterized by the parameters determined from the Fe data.  Abundances of a given species were normalized  
on a line-by-line basis using the values derived from the solar spectrum, and then averaged together.  Typical
errors in the mean are only 0.01-0.02 dex, indicative of the quality of the data.  The sensitivity of the derived
abundances to arbitrarily selected fiducial variations in the stellar parameters (${\pm}100$ K in $T_{\rm eff}$; 
${\pm}0.2$ dex in log $g$; and ${\pm}0.2$ km s$^{-1}$ in microturbulence) are provided for each element in Table 3.  
Coupling these with the parameter uncertainties and the statistical uncertainties in the mean yielded total uncertainties 
in the abundance ratio of each element.  The mean abundances and the $1{\sigma}$ uncertainties are given in Table 4. 
\marginpar{Tab.~3}
\marginpar{Tab.~4}

\subsection{Oxygen Abundances}

O abundances were derived from the measured equivalent widths of the ${\lambda}6300$ [O I] feature (Table 2) using the 
{\sf blends} package in {\sf MOOG} to account for contamination by a \ion{Ni}{1} feature at 6300.34.  Isotopic 
components \citep{Jo2003} of Ni were taken into account with the $gf$ values taken from \citet{BFL}; the [O I] $gf$ value 
(-9.717) is taken from \citet{APLA}.  The assumed Ni abundances were taken as [Ni/H]=0.00, -0.04, -0.16, and -0.01 for HIP 
71813, 76114, 77718, and 78399 respectively.  Abundances are given in Table 4.  Uncertainties in [O/H] are dominated by 
those in the equivalent widths ($0.5$ m{\AA}) measurements of the stars and the Sun, and that in log $g$ ($0.12$ 
dex).  These uncertainties from these 3 sources were added in quadrature to yield the total uncertainties associated with
the [O/H] values given in Table 4.   

\subsection{Lithium Abundances}

Li abundances were derived from the ${\lambda}6707$ \ion{Li}{1} resonance features via spectrum synthesis.  Utilizing the 
derived parameters, synthetic spectra of varying Li abundance were created in {\sf MOOG} using the line list from
\citet{King97}.  No contribution from $^6$Li was assumed, a reasonable assumption given that the Li abundances in our
objects are well-below meteoritic (log $N$(Li)=3.31; $^6$Li/$^7$Li$=0.08$).  Smoothing was carried out by convolving the 
synthetic spectra with Gaussians having FWHM values measured from clean, weak lines measured in our spectra.  Comparison 
of the syntheses (solid lines) and the Keck/HIRES spectra in the ${\lambda}6707$ region are shown in Figures 3 and 4.  
Total uncertainties include those due to uncertainties in the $T_{\rm eff}$ value (Table 3) and in the fit itself.  The Li 
results are listed in Table 4.   
\marginpar{Fig.~3}
\marginpar{Fig.~4} 

\subsection{Rotational Velocity and Chromospheric Emission}

The same FWHM values measured for each star and used to smooth the syntheses were assumed to be the quadrature sum of 
components due to spectrograph resolution and (twice the projected) rotational velocity.  The resulting $v$ sin $i$ values are
listed in Table 4.  Inasmuch as we assume no contribution from macroturbulent broadening mechanisms, we present these 
estimates as upper limits to the projected rotational velocity.  The \ion{Ca}{2} H\&K chromospheric emission indices of our 
objects are listed in Table 4 and come from the low-resolution ($R{\sim}2000$) KPNO coude' feed-based survey of D.~Soderblom.

\subsection{Masses and Ages}

Masses and ages of the Sun and our four solar twin candidates were estimated by placing them in the $M_V$ versus $T_{\rm eff}$ 
plane using our temperature estimates and the Hipparcos-based absolute visual magnitudes.  Comparison of these positions
with isochrones and sequences of constant mass taken from appropriate metallicity Yonsei-Yale Isochrones \citep{Yi2003} (as 
updated by \citet{Dem2004}) yielded the mass and age estimates in Table 4.  The uncertainties in mass and age are calculated
assuming the influence of uncertainties in our $T_{\rm eff}$ and $M_V$ values; including the uncertainty in our metallicity estimates 
(${\sigma}{\sim}0.04$ dex) has a negligible effect on the uncertainty of our estimated masses, but would contribute an additional 
0.4 Gyr uncertainty in the age estimates.  The HR diagrams containing our objects and these isochrones are shown in Figure 5.  
\marginpar{Fig.~5} 
 
\section{Results and Discussion}

\subsection{Comparison with Previous Results}

HIP 71813 is included in the recent Geneva-Copenhagen solar neighborhood survey of \citet{Nord04}.  Their photometric 
metallicity determination of [Fe/H]$=+0.01$ is in outstanding agreement with our Al, Ca, Ti, Fe, and Ni abundances, which 
range from $-0.02$ to $+0.02$.  Their photometric $T_{\rm eff}$ estimate of 5662 K is some 90 K lower than our 
spectroscopic value.  If the solar color, $(B-V)_{\odot}=0.642$, adopted in Table 4 is to be believed, then our $T_{\rm eff}$ 
value would seem to be more consistent with the nearly indistinguishable $(B-V)$ index (0.644) of HIP 71813. 

HIP 76114 is also included in the Geneva-Copenhagen survey.  The \citet{Nord04} photometric metallicity
of [Fe/H]$=-0.05$ is also in outstanding agreement with our Al, Ca, Ti, Fe, and Ni abundances, which range from
$-0.06$ to $-0.02$.  The photometric $T_{\rm eff}$ difference between HIP 71813 and 76114 (former minus latter) 
of 52 K is in excellent agreement with our spectroscopic difference of 40 K.     

HIP 77718 has a photometric metallicity, [Fe/H]$=-0.19$ from the \citet{Nord04} solar neighborhood survey
that is in good agreement with our Al, Ti, Fe, and Ni determinations, which range from $-0.15$ to $-0.22$; our
[Ca/H] abundance of $-0.09$ appears only mildly anomalous in comparison.  The HIP 77718 minus 71813 photometric 
$T_{\rm eff}$ difference of 92 K is in outstanding agreement with the 90 K spectroscopic difference.  \citet{Gray2003} 
have determined the parameters and overall abundance of HIP 77718 via the analysis of low resolution blue spectra as 
part of their NStar survey.  The independent spectroscopic $T_{\rm eff}$ estimate, made via different comparisons of 
different spectral features in a different part of the spectrum, of 5859 K is only 19 K larger than our own and 105 K 
larger than the photometric value.  The \citet{Gray2003} metallicity of [m/H]$=-0.15$ is indistinguishable from our own 
result.   

HIP 78399 has not been subjected to any published abundance or high-resolution spectroscopic analysis that we are 
aware of.  Accordingly, it lacks a radial velocity determination.  We remedied this by determining a radial 
velocity relative to HIP 76114 via cross-correlation of the spectra in the 6160-6173 {\AA} range.  We assumed the
precision radial velocity of $-35.7$ km/s from \citet{Nid2002} for HIP 76114.  Cross-correlation of the telluric B-Band
spectra in the 6880 {\AA} region revealed a 9.9 km/s offset between the spectra.  While larger than anticipated,
this intra-night drift was confirmed by comparison of telluric water vapor features in the 6300 {\AA} region.  
Accounting for this drift and the appropriate relative heliocentric corrections, we find a radial velocity
of $-24.7{\pm}0.7$ km/s for HIP 78399. 

\subsection{HIP 71813 and the Wolf 630 Moving Group}

\citet{Egg69} included HIP71813 as a member of the Wolf 630 moving group.  Membership in this putative kinematic 
population was defined by Eggen in a vast series papers as traced in the work of \citet{McDH83}.  Regardless
of one's view on the reality of these kinematic assemblages, it is likely that the recent passing of O.~Eggen 
has meant that a wealth of modern data (in particular {\it Hipparcos\/} parallaxes and precision radial velocities) 
has not yet been brought to bear on the reality, properties, and detailed membership of the Wolf 630 group.  A notable 
exception is the work of \citet{SHC}, who find a clustering of late-type stars at $(U,V)=(+20,-30)$ km s$^{-1}$ that is 
absent in the kinematic phase space of their early-type stellar sample; this is highly suggestive of an old moving group 
at Eggen's suggested position of the Wolf 630 group in the Bottlinger ($U$,$V$) diagram.  The salient characteristics 
identified by Eggen for the Wolf 630 group are a) a kinematically old disk population, b) a characteristic Galactic rotational 
velocity of $V=-33$ km/s, and c) a color-luminosity array similar to M 67.   
 
It is beyond the scope of this paper to revisit or refine characteristics of the Wolf 630 group.  However, several 
notes can be made.  First, our 8 Gyr age estimate for HIP 71813 is certainly consistent with an old disk object.  Second, 
if the estimate of \citet{Tay2000} of [Fe/H]$=-0.12$ for the Wolf 630 group metallicity is accurate, then HIP 71813 would 
not seem to be a member.  Third, using {\it Hipparcos\/} parallaxes and proper motions, and modern radial velocity 
determinations \citep{Nord04,TR98}, the $UVW$ kinematics of HIP 71813 can be compared with those of Wolf 629, a Wolf 630
group defining member according to Eggen.  The heliocentric Galactic velocities of all our objects are listed in 
Table 4.  The (U,V)=(+21.3${\pm}1.5$,-36.3${\pm}1.3$) results for HIP 71813 are in excellent agreement with 
those for Wolf 629 (+21.0${\pm}1.3$,-33.4${\pm}1.0$), and consistent with the canonical Wolf 630 group values (26, -33) given
by \citet{Egg69}.  None of our other candidate solar twins has kinematics, which are listed in Table 4, consistent with 
those of the Wolf 630 group.  

\subsection{Solar Twin Status Evaluation}

{\it HIP 71813}. The $T_{\rm eff}$ value, light metal-abundances, and chromospheric \ion{Ca}{2} emission of HIP 71813 
are indistinguishable from solar values.  The Li abundance, however, appears to be depleted by a factor of ${\ge}2$ 
compared to the Sun.  More importantly, however, the star appears significantly more evolved than the sun.  The $M_V$ 
and log $g$ values are significantly lower than the solar values, and our estimated age is a factor of 2 older than 
the Sun's.  While clearly an inappropriate solar twin candidate, the star would appear to be an excellent solar analog 
of significantly older age.   

{\it HIP 76114}.  HIP 76114 is marginally cooler than the Sun, ${\Delta}T_{\rm eff}=-67$ K.  While any of the light element 
abundances alone are indistinguishable from solar, taken together they suggest a metallicity some 0.04 dex lower than solar; 
this is confirmed by the photometric metallicity of \citet{Nord04}.  The \ion{Ca}{2} emission and Li abundance is solar
within the uncertainties, but the star appears marginally evolved relative to the Sun as indicated by its slightly
lower $M_V$ and log $g$ values; table 4 suggests that HIP 76114 is ${\ge}1.5$ Gyr older than the Sun.  This object
may be a suitable solar analog of slightly older age, albeit of likely slightly lower metallicity, that can be included 
in studies looking at solar evolution.   

{\it HIP 77718}.  While the \ion{Ca}{2} chromospheric emission and age determination of HIP 77718 are observationally
indistinguishable from the Sun, our analysis indicates this star is clearly warmer (${\Delta}T_{\rm eff}=63$ K) 
and some 0.16 dex metal-poor relative to solar; both the warmer temperature and slightly metal-poor nature are 
independently confirmed by the spectroscopic analysis of \citet{Gray2003}.  The Li abundance is some
20 times higher than solar.  This difference may be related to reduced PMS Li depletion due to lower metallicity or 
reduced main-sequence depletion due to a younger age; our observations can not distinguish between these 
possibilities.  Regardless, this star is not a good solar twin candidate, nor an optimal metal-poor or younger
solar analog.   

{\it HIP 78399}.  The poorly-studied HIP 78399 appears to hold great promise as a solar twin candidate.  Its 
$T_{\rm eff}$, luminosity, mass, age, light metal abundances, and rotational velocity are all indistinguishable from 
solar values.  The only marked difference seen is the Li abundance, which is a factor of ${\sim}6$ larger than
the solar photospheric abundance.  While the evolution of Li depletion in solar-type stars is a complex and still 
incompletely understood process subject to vigorous investigation, this difference may suggest a slightly younger 
age for HIP 78399, which is allowed by our age determination and may be consistent with a slightly larger \ion{Ca}{2} 
chromospheric flux. 

Currently, the ``closest ever solar twin'' title belongs to HR 6060 \citep{DMDS}.  Several spectroscopic 
analyses of this star have been carried out \citep{LH2005,AP2004,Gray2003,DMDS}.  $T_{\rm eff}$ estimates
range from 5693 to 5835 K, and [Fe/H] estimates from --0.06 to +0.05; the precision $T_{\rm eff}$ analysis
using line ratios \citep{Gray1995} indicates a $T_{\rm eff}$ difference with respect to the Sun of 
17 K.  The \ion{Ca}{2} H\&K emission index (-5.00, Gray 1995) and rotational velocity (${\le}3$ km/s, 
De Mello \& Da Silva 1997) are indistinguishable from solar values.  The {\it Hipparcos\/}-based 
absolute magnitude strongly suggests the mass and age of HR 6060 are virtually identical to the Sun's \citep{DMDS}. 
Just as for HIP 78399, the only glaring outlying parameter is Li abundance, which is a factor of ${\sim}4$ 
larger than the solar photospheric Li abundance \citep{S97}.  The work of \citet{JFS} on 1 $M_{\odot}$ stars
in the solar-age and -abundance cluster M67 suggests that we can expect such objects to exhibit a ${\sim}1$ dex
range in Li; thus, the Sun may not be an especially good Li "standard".  

Based on our analysis, we believe there is a case to be made that HIP 78399 share the stage with
HR 6060 as the closest ever solar twin.  For those engaged in studies of solar twins or the Sun in time, 
HIP 78399 is certainly worthy of closer follow-up study.  Particularly valuable would be:  a) refining
its T$_{\rm eff}$ and luminosity estimates relative to the Sun via Balmer line profile fitting, analysis
of line ratios, etc.~{\ }b) analysis of the ${\lambda}7774$ \ion{O}{1} lines to confirm whether its O abundance 
is truly subsolar, c) performing an independent check on its relative age via the [Th/Nd] ratio \citep{MKB}, and d) 
determining a $^9$Be abundance, which is more immune to the effects of stellar depletion and also contains embedded
information about the ``personal'' integrated Galactic cosmic-ray history of matter comprised by the star. 

\section{Summary}

We have carried out high S/N high-resolution Keck/HIRES spectroscopy of four candidate solar twins drawn from a 
{\it Hipparcos\/}-defined \ion{Ca}{2} H\&K survey.  Parameters, abundances, masses, ages, and kinematics have
been derived in a differential fine analysis.  Comparisons suggests that the {\it relative\/} photometric $T_{\rm eff}$ 
values of \citet{Nord04} and our spectroscopic temperatures are indistinguishably robust; however, the photometric
$T_{\rm eff}$ values are typically 100 K cooler.  There are several lines of evidence that suggest the photometric
scale is misanchored (at least near solar $T_{\rm eff}$).  First, if the solar color of \citet{CdS} is nearly
correct, then our spectroscopic $T_{\rm eff}$ values are in outstanding accord with the colors of HIP 71813 and 78399. 
Second, the independent analysis of HIP 77718 by \citet{Gray2003} using different spectral features in the blue 
yields a spectroscopic $T_{\rm eff}$ in outstanding agreement with our own.  Third, the \citet{Nord04} photometric
$T_{\rm eff}$ estimate for the ``closest ever solar twin'' HR 6060 is 5688 K, some $100$ K lower than the precision
$T_{\rm eff}$ estimate of \citet{Gray1995}.    

At the same time, our light metal-abundances are in excellent agreement with the photometric metallicity estimates for 
the 3 of our objects in \citet{Nord04}, differing by no more than a few hundredths of a dex.  There is no sign of the 
abnormally low photometric metallicity values seen for some very cool Pop I dwarfs in the Hyades and UMa group as noted 
by \citet{KS05}. As these authors note, anomalous photometric estimates may be restricted to late G dwarfs.  Our 
spectroscopic metallicity for HIP 77718 is in nearly exact agreement with that derived from low-resolution blue spectra
by \citet{Gray2003}.      

We present the first abundances and radial velocity estimate for HIP 78399.  Using the radial velocities and 
{\it Hipparcos\/} proper motions and parallaxes, we derive the $UVW$ kinematics of our four solar twin candidates.  
The position of HIP 71813 in the $(U,V)$ plane is consistent with membership in Eggen's Wolf 630 moving group, 
a kinematic structure of late-type {\it Hipparcos\/} stars apparently verified by \citet{SHC}.  Our metallicity
for HIP 71813, [Fe/H]$=-0.02$, is 0.1 dex higher than the Wolf 630 estimate of \citet{Tay2000}, however.  Revisiting
the characteristic metallicity via identification of assured Wolf 630 group members using {\it Hipparcos\/} data and 
new precision radial velocities, and follow-up high resolution spectroscopy to determine abundances would be of great 
value.   

HIP 77718 is ${\sim}70$ K warmer than the Sun, significantly more metal-poor ([m/H]${\sim}-0.16$), significantly 
more Li-rich (log $N$(Li)${\sim}2.3$) and a few percent lass massive than the Sun; we deem it neither a suitable 
solar twin nor solar analog to trace the evolution of the Sun.  The light-metal and Li abundances of HIP 76114
are much closer to solar.  However, HIP 76114 does appear to be slightly metal-poor ([m/H]$=-0.04$), cooler
${\Delta}T_{\rm eff}=67$ K, older ${\Delta}{\tau}{\ge}3$ Gyr, and a few percent less massive compared to the 
Sun.   

HIP 71813 appears to be an excellent solar analog of solar abundance, mass, and $T_{\rm eff}$, but advanced age--$M_V=4.45$ and 
${\tau}{\sim}8$ Gyr; the more evolved state of this star is likely reflected in the subsolar upper limit to its
Li abundance.  Finally, our first ever analysis of HIP 78399 suggests this object may be a solar twin candidate of 
quality comparable to the ``closest ever solar twin'' HR 6060 \citep{DMDS}.  The $T_{\rm eff}$, mass, age, and light
metal abundances of this object are indistinguishable from solar given the uncertainties.  The only obvious difference
is that which characterizes HR 6060 as well-- a Li abundance a factor of a few larger than the solar photospheric value. 
This object merits additional study as a solar twin to refine its parameters; of particular interest will be confirming
our subsolar O abundance derived from the very weak ${\lambda}6300$ [\ion{O}{1}] feature.   

\acknowledgments
We are indebted to Dr.~David Soderblom for the use of his nearby star activity catalog from which our objects
were selected and for his valuable comments on the manuscript.  It is a pleasure to thank Dr.~M.~Novicki for 
her careful and cheerful help in the Keck data reduction.  JRK gratefully acknowledges support for this work from 
NSF awards AST-0086576 and AST-0239518, and a generous grant from the Charles Curry Foundation to Clemson University.  
Additional support was provided by NSF award AST-0097955 to AMB.  SCS was supported by a graduate scholarship from 
the South Carolina Space Grant Consortium.

%%Fig 1
\begin{figure}
\plotone{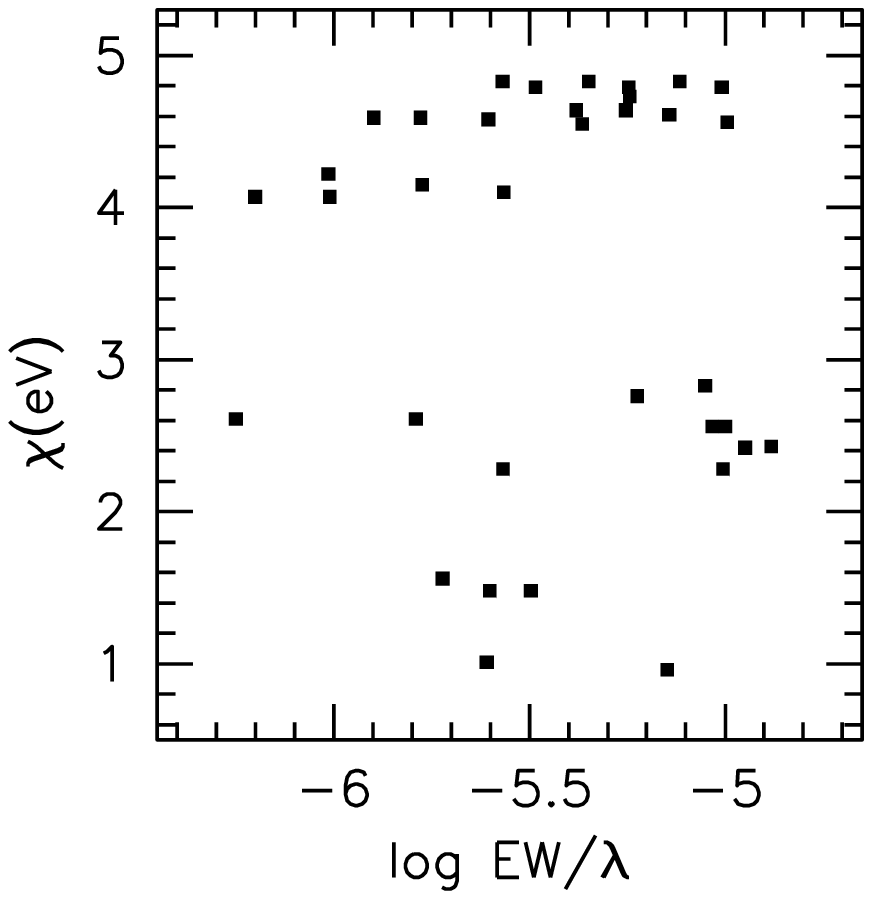}
\caption[]{Lower excitation potential is plotted versus the reduced equivalent width (measured from our
solar proxy spectrum) for the \ion{Fe}{1} lines in our sample.}
\end{figure}

%%Fig 2
\begin{figure}
\plotone{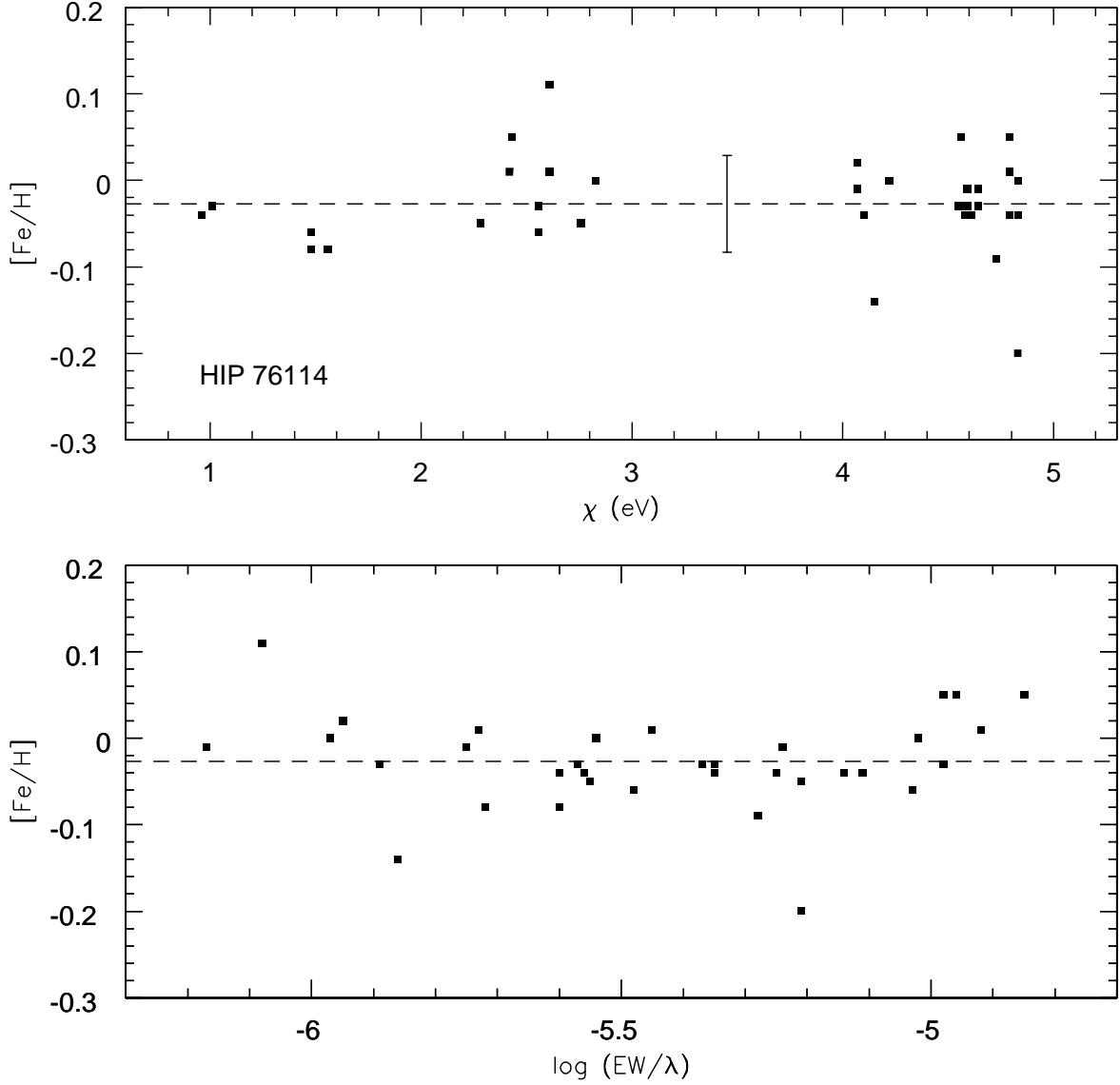}
\caption[]{(Top) The HIP 76114 line-by-line \ion{Fe}{1}-based [Fe/H] abundances with our final model atmosphere 
parameters are shown versus lower excitation potential.  The error bar shows the line-to-line scatter (not the
mean uncertainty). (Bottom) The same [Fe/H] abundances are shown versus reduced equivalent width.  The linear 
correlation coefficients in both panel are ${\sim}0.00$.} 
\end{figure}

%%Fig 3 
\begin{figure}
\plotone{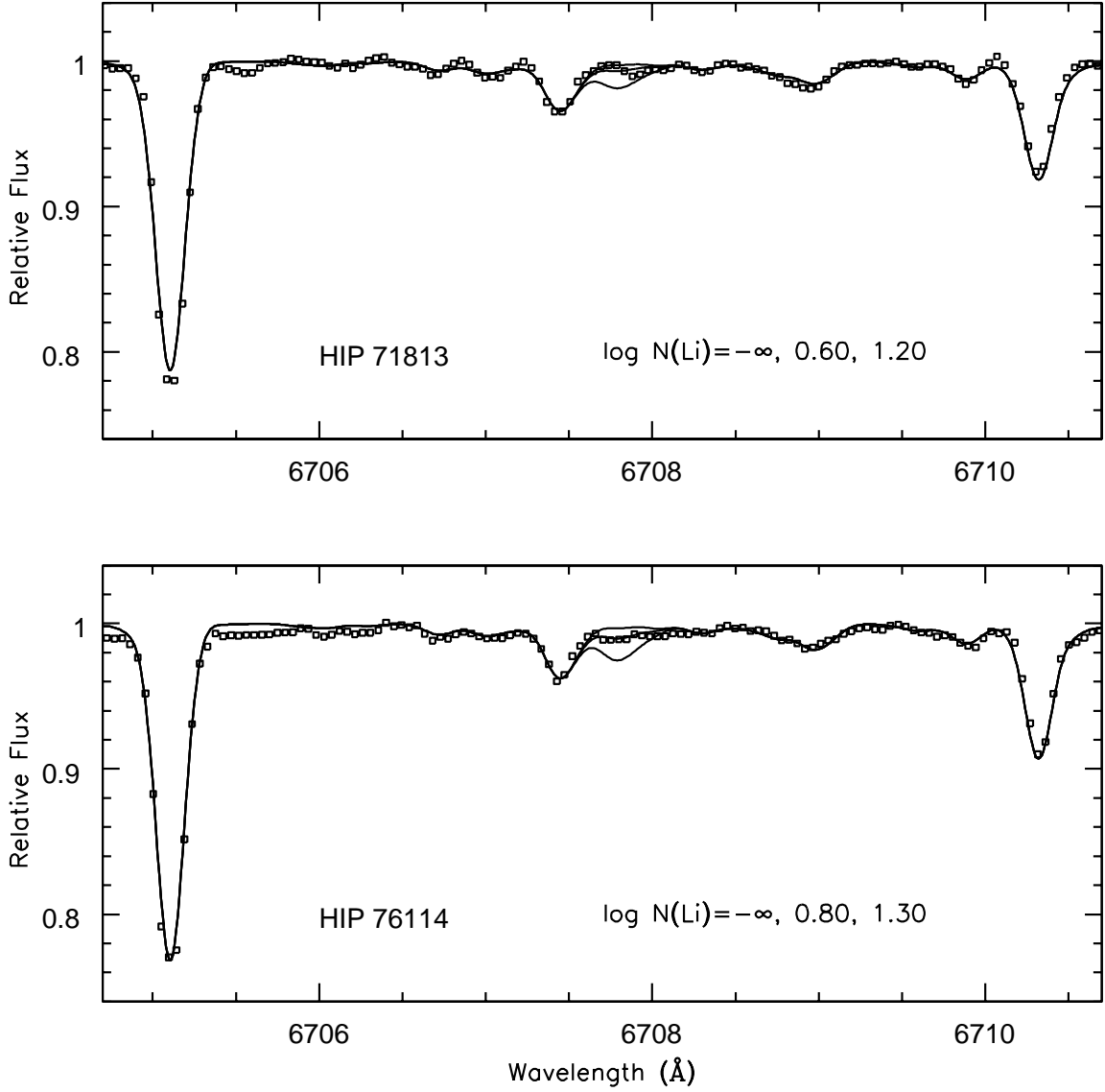}
\caption[]{The ${\lambda}6707$ \ion{Li}{1} region Keck/HIRES spectra (open squares) of HIP 71813 (top) and HIP 76114 (bottom) 
are shown with syntheses of varying Li abundance (solid lines).}
\end{figure}

%%Fig 4 
\begin{figure}
\plotone{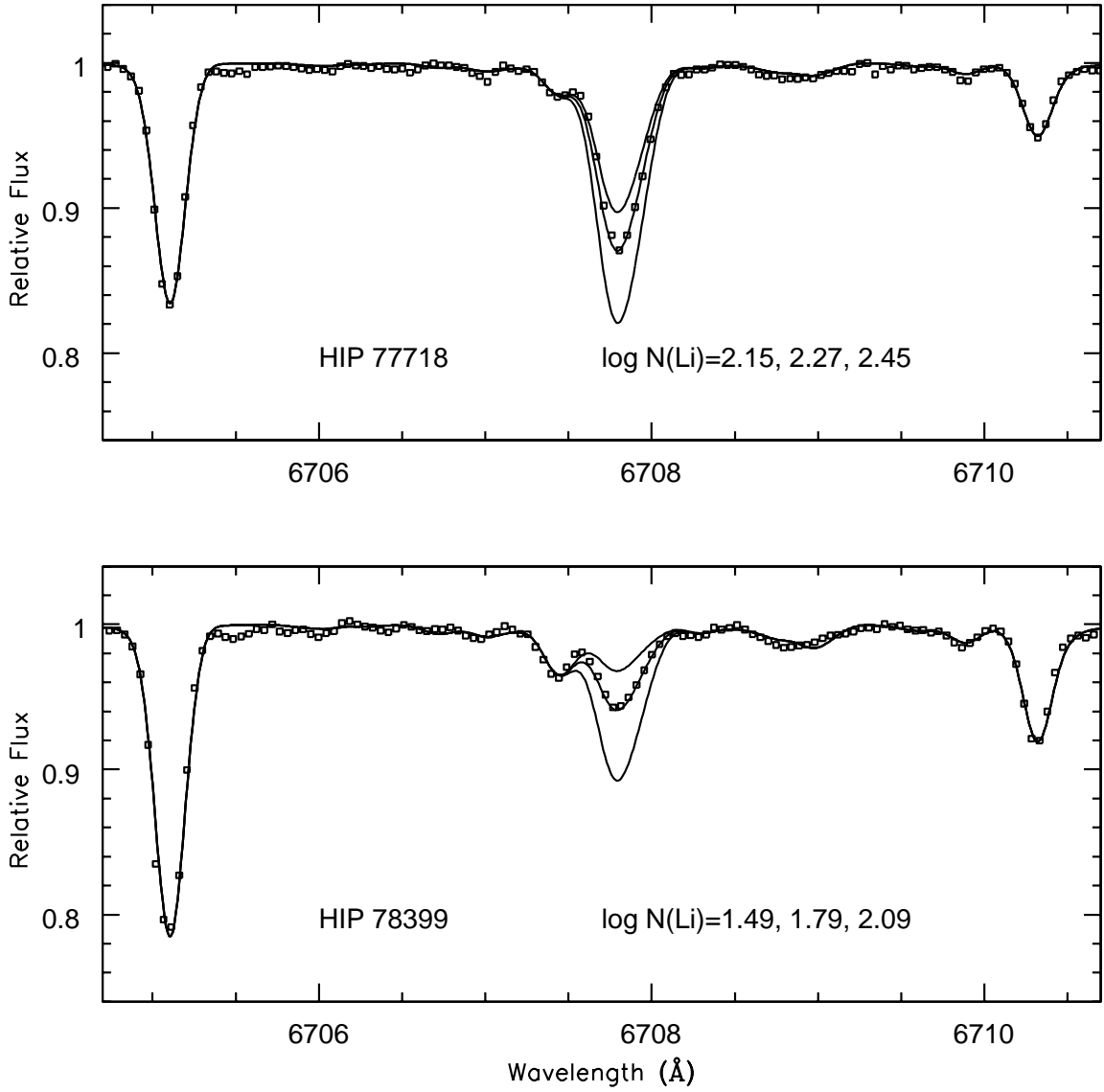}
\caption[]{Same as Figure 2 for HIP 77718 (top) and HIP 78399 (bottom).}
\end{figure}

%%Fig 5 
\begin{figure}
\plotone{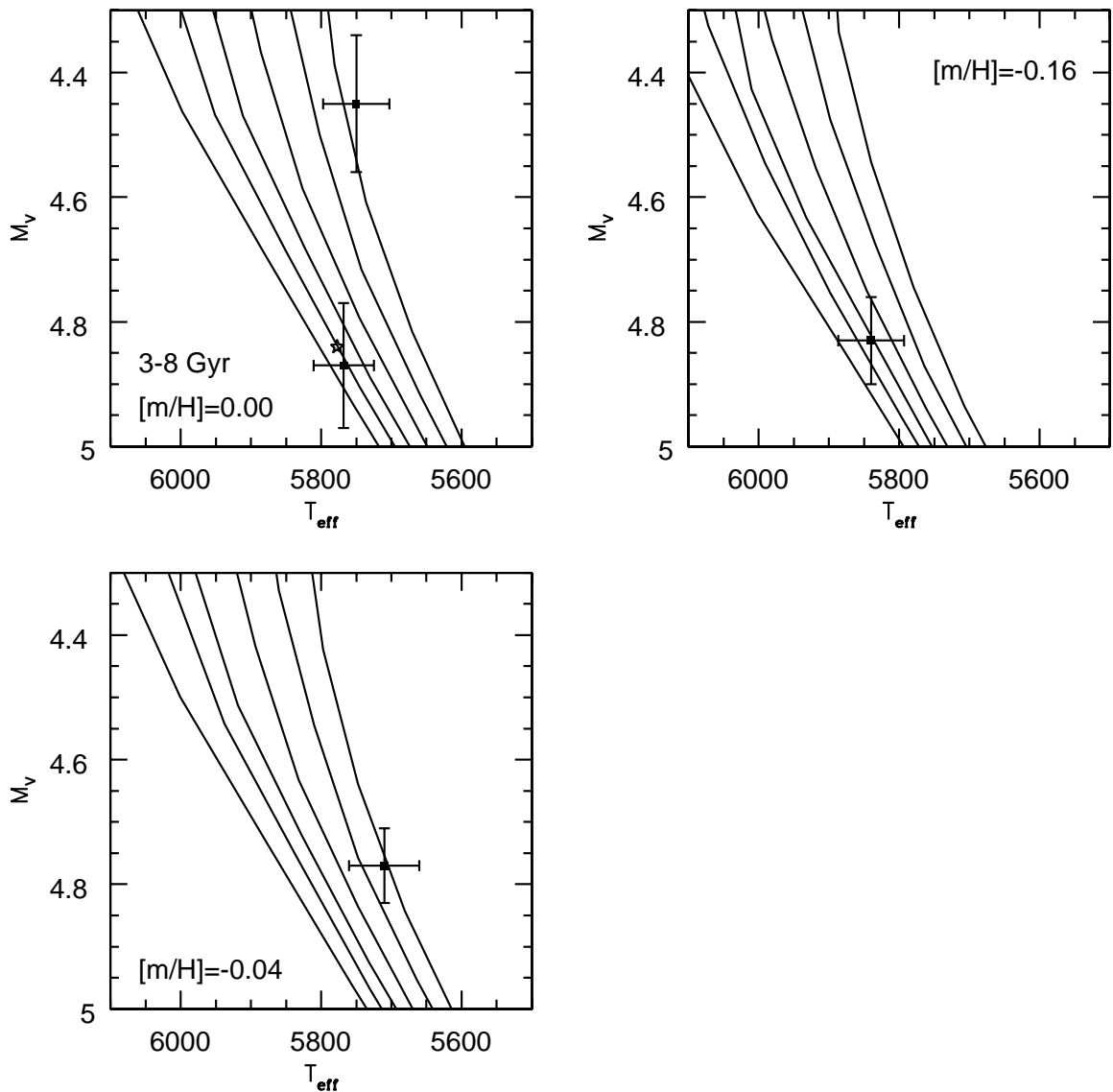}
\caption[]{The Yale-Yonsei evolutionary tracks for 3-8 Gyr for [Fe/H]=0.00 (upper left), -0.04 (lower left), and -0.16 (upper right)
are plotted with our 4 candidate solar analogs, whose locations are defined by our spectroscopic temperatures and the 
{\it Hipparcos\/}-based absolute magnitudes.  The [m/H]=0.00 (upper left) plane contains HIP 71813 and 78399 and the Sun (open star). 
The [m/H]=-0.16 (upper right) plane contains HIP 77718.  The [m/H]=-0.04 plane (lower left) plane contains HIP 76114.}   
\end{figure}

%%TABLE 1
%\documentclass{aastex}
%\begin{document}
\begin{deluxetable}{llrcr} 
\tablecolumns{5}
\tablewidth{0pc}
\tablenum{1}
\tablecaption{Observational Log}
\tablehead{
\colhead{HIP \#} & \colhead{HD \#} & \colhead{$V$} & \colhead{exp.~(s)} & \colhead{S/N} \\ 
}
\startdata
71813 & 129357 & 7.82 & 382 & 456 \\    
76114 & 138573 & 7.23 & 180 & 381 \\  
77718 & 142093 & 7.32 & 180 & 383 \\ 
78399 & 143436 & 8.06 & 300 & 363 \\
\enddata
\end{deluxetable} 

%%TABLE 2 
%\documentclass{aastex}
%\begin{document}
\begin{deluxetable}{lrrrrrrrrrrrrr}
\tabletypesize{\scriptsize}
\rotate
\tablecolumns{14}
\tablewidth{0pc}
\tablenum{2}
\tablecaption{Line Data} 
\tablehead{
\colhead{Species} & \colhead{$\lambda$} & \colhead{$\chi$} & \colhead{log $gf$} & \colhead{EW(Sun)} & \colhead{log N} & \colhead{EW(HIP71813} & \colhead{log N} & \colhead{EW(HIP76114)} & \colhead{log N} & \colhead{EW(HIP77718)} & \colhead{log N} & \colhead{EW(HIP78399)} & \colhead{log N} \\
\colhead{ } & \colhead{\AA} & \colhead{eV} & \colhead{ } & \colhead{m{\AA}} & \colhead{ } & \colhead{m{\AA}} & \colhead{ } & \colhead{m{\AA}} & \colhead{ } & \colhead{m{\AA}} & \colhead{ } 
}
\startdata
\ion{O}{1}  & 6300.3  & 0.00 & -9.72   & 5.5\tablenotemark{a} & 8.69 & 7.3 & 8.72 & 7.5 & 8.76 & 2.8 & 8.18 & 4.8 & 8.47 \\ 
\ion{Al}{1} & 6696.03 & 3.14 & -1.65   & 40.3 & 6.61 & 41.1 & 6.62 & 39.8 & 6.57 & 25.8 & 6.38 & 36.8 & 6.55 \\    
            & 6698.67 & 3.14 & -1.95   & 23.2 & 6.58 & 22.7 & 6.56 & 23.0 & 6.54 & 15.4 & 6.40 & 20.2 & 6.50 \\ 
\ion{Ca}{1} & 6455.60 & 2.52 & -1.50   & 59.7 & 6.47 & 60.2 & 6.50 & 59.5 & 6.43 & 48.9 & 6.36 & 59.1 & 6.46 \\
            & 6464.68 & 2.52 & -2.53   & 13.3 & 6.53 & 14.1 & 6.54 & 13.9 & 6.51 & 10.1 & 6.44 & 13.6 & 6.53 \\ 
            & 6499.65 & 2.52 & -1.00   & 89.8 & 6.45 & 87.4 & 6.46 & 91.5 & 6.45 & 79.7 & 6.38 & 94.6 & 6.53 \\ 
\ion{Ti}{1} & 6599.11 & 0.90 & -2.06   & 10.2 & 5.01 & 11.1 & 5.03 & 10.5 & 4.95 &  6.3 & 4.85 & 10.8 & 5.03 \\ 
\ion{Fe}{1} & 6475.63 & 2.56 & -2.97   & 60.1 & 7.63 & 54.4 & 7.51 & 61.0 & 7.57 & 45.6 & 7.44 & 59.5 & 7.60 \\
	    & 6481.88 & 2.28 & -3.01   & 64.1 & 7.47 & 63.9 & 7.46 & ...  & ...  & 53.5 & 7.36 & 64.2 & 7.45 \\
	    & 6494.50 & 4.73 & -1.44   & 37.1 & 7.78 & 32.3 & 7.67 & 34.4 & 7.69 & 24.2 & 7.54 & 35.4 & 7.74 \\
	    & 6495.74 & 4.83 & -1.11   & 49.8 & 7.77 & 37.8 & 7.55 & 40.3 & 7.57 & 30.5 & 7.45 & 42.5 & 7.63 \\
	    & 6496.47 & 4.79 & -0.65   & 63.7 & 7.51 & 60.5 & 7.47 & 68.7 & 7.56 & 55.3 & 7.42 & 64.0 & 7.51 \\
	    & 6498.94 & 0.96 & -4.70   & 46.2 & 7.49 & 45.4 & 7.44 & 49.9 & 7.45 & 34.0 & 7.34 & 47.6 & 7.49 \\
	    & 6509.61 & 4.07 & -2.97   &  4.1 & 7.51 &  4.5 & 7.53 &  4.4 & 7.50 &  2.7 & 7.36 &  4.3 & 7.52 \\
	    & 6518.37 & 2.83 & -2.67   & 58.0 & 7.56 & 56.7 & 7.52 & 62.0 & 7.56 & 45.9 & 7.41 & 60.7 & 7.59 \\
	    & 6581.22 & 1.48 & -4.82   & 21.0 & 7.61 & 21.7 & 7.59 & 21.8 & 7.55 & 12.9 & 7.41 & 19.6 & 7.55 \\
	    & 6591.33 & 4.59 & -2.07   & 11.0 & 7.57 &  9.7 & 7.49 & 11.7 & 7.56 &  8.6 & 7.49 & 10.1 & 7.52 \\
	    & 6593.88 & 2.43 & -2.34   & 86.6 & 7.36 & 87.3 & 7.39 & 94.2 & 7.41 & 76.8 & 7.29 & 90.3 & 7.41 \\
	    & 6608.04 & 2.28 & -4.02   & 17.9 & 7.52 & 17.6 & 7.48 & 18.7 & 7.47 & 11.9 & 7.36 & 18.0 & 7.50 \\
	    & 6609.12 & 2.56 & -2.67   & 66.2 & 7.43 & 69.1 & 7.49 & 68.7 & 7.40 & 55.8 & 7.32 & 69.2 & 7.47 \\
	    & 6625.04 & 1.01 & -5.38   & 16.3 & 7.55 & 15.8 & 7.50 & 17.9 & 7.52 &  9.7 & 7.36 & 16.8 & 7.55 \\
	    & 6627.56 & 4.55 & -1.59   & 28.6 & 7.58 & 28.9 & 7.57 & 29.3 & 7.55 & 20.2 & 7.41 & 29.5 & 7.59 \\
	    & 6633.43 & 4.83 & -1.35   & 29.8 & 7.63 & 30.2 & 7.63 & 29.9 & 7.59 & 21.2 & 7.46 & 31.1 & 7.65 \\
	    & 6633.76 & 4.56 & -0.79   & 67.2 & 7.49 & 64.7 & 7.47 & 72.1 & 7.54 & 58.5 & 7.41 & 71.7 & 7.57 \\
	    & 6634.12 & 4.79 & -1.32   & 37.7 & 7.72 & 37.4 & 7.71 & 37.3 & 7.68 & 26.8 & 7.53 & 36.6 & 7.69 \\
	    & 6646.97 & 2.61 & -4.01   & 10.8 & 7.57 &  9.7 & 7.49 & 12.4 & 7.58 &  6.2 & 7.36 & 10.1 & 7.52 \\
	    & 6653.91 & 4.15 & -2.53   & 11.2 & 7.61 &  9.7 & 7.52 &  9.2 & 7.47 &  6.8 & 7.41 &  8.5 & 7.47 \\
	    & 6696.32 & 4.83 & -1.65   & 18.1 & 7.64 & 20.4 & 7.69 & 19.4 & 7.64 & 13.4 & 7.51 & 17.6 & 7.61 \\
	    & 6699.14 & 4.59 & -2.22   &  8.5 & 7.59 &  8.7 & 7.59 &  8.6 & 7.56 &  6.1 & 7.47 &  8.7 & 7.59 \\
            & 6703.58 & 2.76 & -3.13   & 40.0 & 7.60 & 39.2 & 7.57 & 41.0 & 7.55 & 28.7 & 7.43 & 38.5 & 7.56 \\
	    & 6704.50 & 4.22 & -2.67   &  6.5 & 7.56 &  5.5 & 7.46 &  7.2 & 7.56 &  5.5 & 7.52 &  8.3 & 7.67 \\
	    & 6710.32 & 1.48 & -4.90   & 16.8 & 7.56 & 15.8 & 7.49 & 16.9 & 7.48 & 10.2 & 7.37 & 16.7 & 7.54 \\
	    & 6713.74 & 4.79 & -1.52   & 22.0 & 7.57 & 22.5 & 7.57 & 23.6 & 7.58 & 15.6 & 7.42 & 22.4 & 7.58 \\
	    & 6716.25 & 4.58 & -1.90   & 16.7 & 7.60 & 16.0 & 7.56 & 16.8 & 7.56 & 11.2 & 7.43 & 17.0 & 7.60 \\
            & 6725.36 & 4.10 & -2.30   & 18.3 & 7.59 & 19.5 & 7.61 & 18.5 & 7.55 & 12.3 & 7.42 & 20.4 & 7.64 \\
	    & 6726.67 & 4.61 & -1.12   & 48.5 & 7.54 & 49.2 & 7.56 & 48.4 & 7.50 & 37.3 & 7.38 & 52.4 & 7.61 \\
	    & 6733.15 & 4.64 & -1.52   & 28.1 & 7.58 & 28.4 & 7.57 & 28.4 & 7.55 & 20.5 & 7.43 & 26.7 & 7.54 \\
	    & 6739.52 & 1.56 & -4.98   & 12.8 & 7.58 & 13.4 & 7.57 & 12.9 & 7.50 &  8.1 & 7.42 & 12.7 & 7.56 \\
	    & 6745.98 & 4.07 & -2.74   &  6.6 & 7.49 & 10.6 & 7.70 &  7.6 & 7.51 &  5.1 & 7.41 &  7.9 & 7.56 \\
	    & 6746.98 & 2.61 & -4.35   &  3.8 & 7.42 & ...  & ...  &  5.6 & 7.53 &  2.8 & 7.33 &  4.6 & 7.49 \\
	    & 6750.16 & 2.42 & -2.48   & 75.9 & 7.27 & 80.7 & 7.37 & 80.7 & 7.28 & 64.7 & 7.16 & 79.0 & 7.31 \\
	    & 6752.72 & 4.64 & -1.30   & 37.7 & 7.55 & 38.5 & 7.56 & 39.1 & 7.54 & 27.6 & 7.39 & 38.1 & 7.55 \\
\ion{Fe}{2} & 6239.95 & 3.89 & -3.59   & 14.1 & 7.69 & 15.6 & 7.64 & 13.7 & 7.60 & 11.9 & 7.51 & 16.3 & 7.71 \\    
            & 6247.56 & 3.89 & -2.55   & 54.8 & 7.68 & 56.3 & 7.62 & ...  & ...  & 51.7 & 7.54 & 55.5 & 7.62 \\  
            & 6385.46 & 5.55 & -2.85   &  3.9 & 7.79 &  5.6 & 7.88 &  5.0 & 7.86 &  4.0 & 7.74 &  5.3 & 7.89 \\  
            & 6407.29 & 3.89 & -3.49   & 33.3 & 8.14 & 35.9 & 8.10 & 33.9 & 8.07 & 25.2 & 7.85 & 34.6 & 8.10 \\  
            & 6446.40 & 6.22 & -2.11   &  4.5 & 7.71 &  5.2 & 7.69 &  5.1 & 7.72 &  4.0 & 7.59 &  4.8 & 7.69 \\  
            & 6456.39 & 3.90 & -2.25   & 64.0 & 7.57 & 65.9 & 7.53 & 63.5 & 7.48 & 60.6 & 7.43 & 69.6 & 7.62 \\  
            & 6506.36 & 5.59 & -3.01   &  3.9 & 7.99 &  4.1 & 7.92 &  4.7 & 8.03 &  3.0 & 7.80 &  3.8 & 7.93 \\  
            & 6516.08 & 2.89 & -3.55   & 55.3 & 7.70 & 59.7 & 7.70 & 55.9 & 7.62 & 52.7 & 7.57 & 57.8 & 7.68 \\ 
\ion{Ni}{1} & 6767.78 & 1.83 & -1.89   & 82.3 & 5.95 & 84.7 & 6.00 & 82.7 & 5.87 & 70.4 & 5.82 & 84.0 & 5.95 \\    
            & 6586.32 & 1.95 & -2.95   & 44.4 & 6.44 & 43.3 & 6.38 & 45.7 & 6.38 & 29.6 & 6.19 & 40.7 & 6.34 \\  
            & 6598.61 & 4.23 & -1.02   & 25.9 & 6.35 & 26.6 & 6.35 & 27.4 & 6.34 & 18.1 & 6.17 & 26.9 & 6.36 \\  
            & 6635.14 & 4.42 & -0.87   & 24.5 & 6.35 & 25.8 & 6.36 & 25.1 & 6.32 & 17.6 & 6.18 & 24.0 & 6.32 \\  
            & 6643.64 & 1.68 & -2.01   & 98.4 & 6.23 & 96.1 & 6.20 & 98.3 & 6.13 & 79.9 & 5.99 & 102.2 & 6.27 \\  
            & 6482.81 & 1.93 & -2.97   & 41.3 & 6.38 & 42.0 & 6.36 & 43.1 & 6.34 & 28.8 & 6.18 & 40.5 & 6.34 \\  
            & 6532.88 & 1.93 & -3.47   & 17.0 & 6.32 & 16.3 & 6.26 & 17.1 & 6.25 &  7.8 & 5.97 & 18.0 & 6.33 \\ 
\enddata
\tablenotetext{a}{The ${\lambda}6300$ [O I] equivalent widths for all stars are presumed to contain a contribution 
from a blending \ion{Ni}{1} feature that was accounted for as discussed in the text.} 
\end{deluxetable}                 

%%TABLE 3
%\documentclass{aastex}
%\begin{document}
\begin{deluxetable}{lrrrr}
\tablecolumns{4}
\tablewidth{0pc}
\tablenum{3}
\tablecaption{Abundance Sensitivities}
\tablehead{
\colhead{Species} & \colhead{${\Delta}T_{\rm eff}$} & \colhead{${\Delta}$log $g$} & \colhead{${\Delta}{\xi}$} \\
\colhead{ }       & \colhead{${\pm}100$ K}          & \colhead{${\pm}0.2$ dex}    & \colhead{${\pm}0.2$ km s$^{-1}$}
}
\startdata
\ion{Li}{1}  & ${\pm}0.09$  & ${\pm}0.02$  & ${\mp}0.00$  \\
\ion{O}{1}   & ${\pm}0.02$  & ${\pm}0.09$  & ${\mp}0.00$  \\ 
\ion{Al}{1}  & ${\pm}0.050$ & ${\mp}0.005$ & ${\mp}0.005$ \\   
\ion{Ca}{1}  & ${\pm}0.067$ & ${\mp}0.016$ & ${\mp}0.033$ \\  
\ion{Ti}{1}  & ${\pm}0.11$  & ${\pm}0.00$  & ${\mp}0.00$  \\ 
\ion{Fe}{1}  & ${\pm}0.070$ & ${\pm}0.005$ & ${\mp}0.023$ \\  
\ion{Fe}{2}  & ${\mp}0.041$ & ${\pm}0.073$ & ${\mp}0.025$ \\ 
\ion{Ni}{1}  & ${\pm}0.073$ & ${\pm}0.013$ & ${\mp}0.039$ \\  
\enddata
\end{deluxetable}

%%TABLE 4 
%\documentclass{aastex}
%\begin{document}
\begin{deluxetable}{lrrrrr}
\tablecolumns{6}
\tablewidth{0pc}
\tablenum{4}
\tablecaption{Solar Twin Candidate Summary} 
\tablehead{
\colhead{Parameter} & \colhead{Sun} & \colhead{HIP 71813} & \colhead{HIP 76114} & \colhead{HIP 77718} & \colhead{HIP 78399} \\ 
\colhead{ }         & \colhead{ }   & \colhead{HD 129357} & \colhead{HD 138573} & \colhead{HD 142093} & \colhead{HD 143436} 
}
\startdata
$M_V$              & 4.83${\pm}0.01$   & 4.45${\pm}0.11$     & 4.77${\pm}0.06$   & 4.83${\pm}0.07$   & 4.87${\pm}0.10$ \\
$(B-V)$\tablenotemark{a} & 0.642${\pm}0.004$ & 0.644${\pm}0.002$   & 0.661${\pm}0.005$ & 0.604${\pm}0.007$ & 0.644${\pm}0.001$ \\
$T_{\rm eff}$(K)   & 5777              & 5749${\pm}47$       & 5710${\pm}50$     & 5841${\pm}47$     & 5768${\pm}43$ \\  
${\xi}$ (km/s)     & 1.25              & 1.22${\pm}0.13$     & 1.35${\pm}0.10$   & 1.18${\pm}0.13$   & 1.32${\pm}0.09$ \\  
$[{\rm m/H}]$\tablenotemark{b} & 0.00  & 0.00                & 0.00              & -0.15             & 0.00 \\
log $g$            & 4.44              & 4.16${\pm}0.13$     & 4.20${\pm}0.15$   & 4.33${\pm}0.15$   & 4.28${\pm}0.12$ \\
$[{\rm Fe/H}]$     & 0.                & -0.02${\pm}0.04$    & -0.03${\pm}0.04$  & -0.15${\pm}0.04$  & -0.00${\pm}0.03$ \\  
$[{\rm Ni/H}]$     & 0.                & -0.02${\pm}0.05$    & -0.06${\pm}0.05$  & -0.22${\pm}0.05$  & -0.02${\pm}0.04$ \\  
$[{\rm Ca/H}]$     & 0.                & +0.02${\pm}0.04$    & -0.02${\pm}0.04$  & -0.09${\pm}0.04$  & +0.02${\pm}0.05$ \\  
$[{\rm Ti/H}]$     & 0.                & +0.02${\pm}0.09$    & -0.06${\pm}0.08$  & -0.16${\pm}0.08$  & +0.02${\pm}0.07$ \\  
$[{\rm Al/H}]$     & 0.                & -0.01${\pm}0.06$    & -0.04${\pm}0.05$  & -0.21${\pm}0.04$  & -0.07${\pm}0.04$ \\  
$[{\rm O/H}]$      & 0.                & +0.03${\pm}0.10$    & +0.07${\pm}0.10$  & -0.51${\pm}0.19$  & -0.20${\pm}0.12$ \\ 
log $N$(Li)        & 1.03${\pm}0.04$   & ${\le}0.6{\pm}0.04$ & 0.8${\pm}0.13$    & 2.27${\pm}0.06$   & 1.79${\pm}0.07$ \\  
log $R_{\rm HK}$   & -4.95             & -4.96               & -5.00             & -4.84             & -4.87 \\
$v$ sin $i$ (km/s) & ${\le}2.5$        & ${\le}2.5$          & ${\le}2.1$        & ${\le}2.8$        & ${\le}2.6$ \\  
M (M$_{\odot}$)\tablenotemark{c} & 1.01              & 1.00${\pm}0.06$     & 0.97${\pm}0.015$  & 0.975${\pm}0.015$ & 1.01${\pm}0.02$ \\
Age (Gyr)\tablenotemark{c}       & 4.2${\pm}0.2$     & 8.2${\pm}1.3$       & 7.8${\pm}2.0$     & 5.0${\pm}2.3$     & 3.8${\pm}2.9$  \\ 
$U$ (km/s)         & & +21.3${\pm}1.5$ & -37.2${\pm}0.4$ & -5.6${\pm}0.3$  & -19.2${\pm}0.5$ \\     
$V$ (km/s)         & & -36.3${\pm}1.3$ & +9.0${\pm}0.4$  & -26.3${\pm}0.6$ & -38.6${\pm}1.6$ \\ 
$W$ (km/s)         & & -32.0${\pm}0.4$ & -19.1${\pm}0.3$ & -16.9${\pm}0.2$ & -7.0${\pm}0.5$  \\ 
\enddata
\tablenotetext{a}{The contentious solar color is taken from \citet{CdS}.}
\tablenotetext{b}{The metallicity characterizing the model atmosphere grids used in the abundance analysis.} 
\tablenotetext{c}{Masses and ages derived from comparison of evolutionary tracks and position in the $M_V$ vs.~$T_{\rm eff}$ plane.}
\end{deluxetable}

\end{document}